\documentstyle[twocolumn,prl,aps,epsfig]{revtex}
\renewcommand{\vec}[1]{{\mathbf{#1}}}
\newcommand{\beq}{\begin{eqnarray}} 
\newcommand{\eeq}{\end{eqnarray}} 

\begin{document} 
\draft 
 
\title 
{d$_{x^2-y^2}$ Pairing of Composite Excitations in the 2D Hubbard Model}
\author{Tudor D. Stanescu, Ivar Martin$^*$, and Philip Phillips}

%
\address
{Loomis Laboratory of Physics\\
University of Illinois at Urbana-Champaign\\
1100 W.Green St., Urbana, IL, 61801-3080}

%

\address{\mbox{ }}
\address{\parbox{14.5cm}{\rm \mbox{ }\mbox{ }
We report on a strong coupling approach (on-site Coulomb
repulsion, $U$ larger than the nearest-neighbour hopping energy $|t|$) to the 
 Hubbard model. Starting from 
the Hubbard operators which diagonalize the interaction term,
we generate a hierarchy of composite operators from the equations
of motion.  Using the Hubbard operators as a basis, we are able to compute
the associated Green functions including
the anomalous Green functions which describe 
pair formation.  We show explicitly
that these anomalous Green functions
are non-zero in the d$_{x^2-y^2}$ channel; however, the entities that pair up
are not single electron-like particles
but rather composite excitations (which we call cexons)  made out of an electron
and a hole on nearest-neighbour sites.  Cexons are fermionic
in nature as they have spin 1/2 and also have unit charge.
Our calculations of 
the chemical potential reveal that 
negative compressibility in the 2D Hubbard model and composite excitation
pairing are intimately
connected, namely, the larger the negative compressibility, the larger
the pairing amplitude.  Our observation of negative compressibility 
in the under-doped regime is consistent with phase segregation
or stripe formation in the normal state.  While pairing ameliorates the
negative compresssibility, it does not eliminate it entirely.
In addition, we find that the anomalous correlation
functions are particle-hole symmetric and exhibit a maximum
at a doping level of roughly 10$\%$ as measured from
half-filling.   For $U=8|t|$, the onset temperature for pair
formation is $0.02|t|$.   
The effect
of nearest-neighbour Coulomb repulsions is discussed.}}
\address{\mbox{ }}
\address{\mbox{ }}

\maketitle

\section{Introduction}
After a protracted quest\cite{history} for superconductivity 
in the Hubbard
model, it is surprising that the question is still open.  This state 
of affairs has arisen primarily because the parameter space in which superconductivity
is anticipated to reside is precisely the strong coupling limit,
$U\gg |t|$, in
which traditional perturbative schemes fail.  As a result, progress on this question
has relied predominantly on exact diagonalization or Quantum
Monte Carlo (QMC) simulations on finite systems.  While early work\cite{history} showed 
promising results on the possible onset of superconducting pair correlations in the 
2D Hubbard model, the most recent numerical work\cite{zcg} shows that true superconductivity
with at least algebraic decay of anomalous correlations is absent.  In light
of such results, it has
also been suggested that some combination of 
phonons\cite{sumit} and or 
next nearest-neighbour hopping\cite{husslin} are needed for
superconductivity to survive.  Further, Kuroki and Aoki\cite{aoki}
have suggested that QMC simulations have yet to access
the energy scale of order $0.01t$ where strong-coupling superconductivity\cite{bs}
is expected to occur.  In addition, Beenen and Edwards\cite{be}
have analyzed the 2D Hubbard model with an equations of motion technique
considerably enhanced by Mancini, Matsumoto, and co-workers~\cite{mmc} and shown 
that pairs with $d_{x^2-y^2}$ symmetry emerge.  However, this
work has key
weaknesses: the onset temperature and the window
of doping over which d$_{x^2-y^2}$ pairing occurs
are highly sensitive to the decoupling scheme of the anomalous Green
functions.  For example, at least order of magnitude variations in the onset
temperature were observed\cite{be} for $U=8|t|$ and a decoupling scheme which 
is supposedly
exact in the limit of $U\rightarrow\infty$ resulted in a near vanishing
of pairing for $U>10|t|$.  

Despite these difficulties, the Hubbard model remains 
central to such strongly correlated problems as superconductivity
in the cuprates as well as the organic conductors.  In fact,
in so far as the motion of holes
in the CuO$_2$ planes in the cuprates can be modeled
 realistically\cite{anderson}
by the Hubbard model in the vicinity of half-filling, the microscopic 
underpinnings
of the pairing mechanism in these materials should arise from this model.  
Because
D=2 is the marginal dimension for superconductivity, it might be that while
the pairing mechanism originates within a single layer,
interlayer
tunneling is needed to maintain long-range phase coherence.  
Hence, the questions
that face the Hubbard model are three-fold: 1) does local pairing of the right
 sort
obtain to describe the cuprates, 2) what entities are involved in the pairing,
and 3) is long-range phase coherence present?  In this work we
develop a strong-coupling approach that is sufficient to answer the first
two questions.  Our approach is based on a simple observation first made
by Hubbard\cite{hubbard}.  Namely, that the bare electron annihilation operator 
can be written as a sum of two composite operators
\beq
c_{i\sigma}=c_{i\sigma}n_{i-\sigma}+c_{i\sigma}(1-n_{i-\sigma})
=\eta_{i\sigma}+\xi_{i\sigma}.
\eeq
The quantities described by $\eta_{i\sigma}$ and $\xi_{i\sigma}$
are composite excitations.  The $\eta$ excitation
describes an electron restricted to move on sites already occupied with
an electron of opposite spin whereas $\xi$ demands that there be no prior
occupancy on the site.  A key feature these operators possess is that
they exactly solve the $t=0$ Hubbard model.
Consequently, an equation of motion approach based on these operators
will lead naturally to an expansion in $t/U$. Such an expansion
is ideally suited for the strong coupling regime $U\gg |t|$ in which
both the cuprates as well as the organic conductors reside. It is this approach
that we pursue here. 

When the kinetic energy is treated as a perturbation,
hybridization is introduced into the Hubbard atomic orbital
basis.  As a result, $\xi_i$ and $\eta_i$ are no longer
the eigen-operators.  The new type of excitations can be thought
of heuristically as resonant valence-bond hybrids.  To determine what
operators create such excitations,
it is sufficient to construct the equations of motion for
the Hubbard operators.  The relevant operators 
have the kinetic energy, $t$ as a coefficient.
As we will see, the
simplest terms that arise are of the form, 
$\eta_{ij\sigma}=c_{i\sigma}n_{j\tau}$ with $(ij)$ nearest
neighbours.  This operator is the generator of a composite
excitation (or cexon) that is restricted to live on sites $i$ with a nearest
neighbour site $j$ occupied. 
Such composites have unit charge as can be seen directly from the 
commutator,
\beq
[\eta_{ij\sigma},n_{l\sigma'}]=\delta_{\sigma,\sigma'}
\delta_{il}\eta_{ij\sigma}.
\eeq
Also, quite obviously, its spin is 1/2.  In fact, 
all the composite excitations generated in our scheme have unit charge.
Hence, spin and charge are coupled.
However, while the $\eta_{ij}$ excitation is fermionic in character, it is not
confined to a single site.  Rather, it is an excitation that lives on 
neighbouring sites. In traditional perturbative
schemes, such as Fermi liquid theory,
 similar composite operators naturally appear. However, in such
approaches, the operators that are generated do not describe new
excitations but rather dress the non-interacting
quasiparticles.  In this approach, the new operators do
in fact describe fundamentally new excitations\cite{zacher}
as they give rise to new peaks in the spectral function.  The appearence
of new peaks in the spectral function is incompatible with a Fermi
liquid description of the composite
excitations.  

It might be argued that the emergence of cexons in this model and the
unrecoverability of Fermi liquid theory
are inevitable in an approach based fundamentally on the atomic
limit of the Hubbard model.  However, any theory of high $T_c$ must be based
on a starting point that captures the essence
of strong-correlation physics.  While the atomic
limit might seem extreme, it is justified
in this context as $U\gg t$ in high $T_c$ systems. Consequently,
it stands to reason that the physically-relevant excitations 
should bear a greater family resemblance
to the atomic limit eigentates than they would to the plane-wave
excitations which populate the non-interacting limit.  Cexons are 
the simplest local excitations that arise once the Hubbard operators
are hybridized on nearest-neighbour sites.  To determine
if such excitations really do constitute new particles in the strongly-coupled
limit, dynamical corrections must be included to the theory we present below.

As in earlier composite operator approaches~\cite{be,mmc,takada,dag1},
we find that the traditional
correlation function, $\langle c_{i\uparrow}c_{j\downarrow}\rangle$, does
{\it not} determine the pairing gap in the 2D Hubbard model.  Rather
\beq\label{thc}
\theta_{ij}=\langle c_{i\uparrow}c_{i\downarrow}n_{j\tau}\rangle=
\langle \eta_{ij\uparrow}\eta_{ij\downarrow}\rangle
\eeq
is the relevant anomalous correlation function that determines pairing in the
d-wave channel.  The emergence of
this correlation function 
as the d-wave
order parameter in the Hubbard model and the
seemingly innocuous rewriting (overlooked previously in similar Green
function analyses\cite{be,mmc}) in terms of 
the $\eta_{ij\sigma}$ operators is particularly illuminating. Pairing
based on the $\eta_{ij\sigma}$ excitations is the new feature which we
develop explicitly in this work.
This rewriting signifies that composite excitations rather than electron-like particles, as would
be the case in a traditional Fermi liquid, are involved in the pairing
process in a strongly-correlated system with repulsive interactions.  
In addition, the fact that the new composite operators describe the pairing
lends further credence to their correspondence with new excitations in
the strongly-correlated regime.
While it is well
known that quasiparticles
in strongly-correlated systems do not resemble well
electron-like excitations, as shown, for example,
by the
absence of a sharp peak at the Fermi surface
in angle-resolved photoemission experiments\cite{arpes} on the
under-doped normal state
of the cuprates, the traditional mechanism invoked to explain
this fact is spin-charge separation\cite{anderson,fisher}.
In 1-dimension\cite{luttinger} spin-charge separation is on firm footing.
In addition, even
in a singlet BCS superconductor, spin-charge separation occurs naturally
as the Cooper pairs carry charge but no spin, whereas the quasiparticles
are spin 1/2 but have no well-defined charge. In D=2, fractionalization
of the electron also occurs in the 
fractional quantum Hall effect\cite{laughlin,exp}.
For strongly-correlated systems, spin-charge separation
has been proposed based on a slave-boson picture\cite{s0,s1,s2,s3,s4,s5,s6,s7,s8}.
However, recently Nayak\cite{nayak} has shown that the slave particles
always remain confined. Our work suggests that rather than falling
apart, electrons form clusters or composites when the Coulomb interaction
is large and repulsive. Such entities form pairs in the 2D Hubbard model.
We propose that composite excitations of the kind we describe here
offer a natural explanation for the absence of electron-like excitations
in the ARPES experiments\cite{arpes} on the underdoped cuprates.  
While it is
tempting to draw a connection between
pairing of composite excitations in the 2D Hubbard
model and the pairing of composite fermions\cite{jain} in
quantum Hall systems, this connection would be tenuous at best.

Our approach then accomplishes two things:  1) First we are able to
access the strong-coupling limit of the Hubbard model and identify
the relevant composite excitations.  2) We are able to
show that such composites pair in the $d_{x^2-y^2}$ channel. 
Our work then represents
a formulation of d-wave pairing from a microscopic model
out of which emerges a hierarchy of new composite fermionic
excitations.  
We find that the magnitude of the anomalous
correlation function for composite excitation
pairing increases as the compressibility in the normal
state becomes more negative.   In the next section,
we outline the essential details of the strong-coupling approach.
In Sec. III, we derive the equations of motion for the composite
excitations and all the relevant correlation functions.  The results 
for the anomalous pairing correlation functions are presented in
Sec. IV.

\section{Formalism}

The starting point of our analysis is the on-site Hubbard model
\beq\label{HHam}
H = -\sum_{i,j,\sigma} t_{ij}c_{i\sigma}^{\dagger}c_{\sigma j} + 
U\sum_{i} n_{i\uparrow}n_{\downarrow} - \mu \sum_{i} n_i
\eeq
where $t_{ij}=t$ if $(i,j)$ are nearest-neighbour sites and zero otherwise,
$\mu$ is the chemical potential and $U$ the on-site Coulomb repulsion.
Consider the retarded Green
function 
\beq
\langle\langle c_{i\uparrow};c_{j\downarrow}\rangle\rangle=
\theta(t-t')\langle \{c_{i\uparrow}(t),c_{j\downarrow}(t')\}\rangle
\eeq
where $\{A,B\}$ is the anticommutator.  
As a result of the interaction term in Eq. (\ref{HHam}), the equation 
of motion for this Green function will generate a new Green function that is proportional
to $\langle\langle [c_{i\uparrow},H];c_{j\downarrow}\rangle\rangle$. This
Green function will contain the composite excitation
$\eta_{i\sigma}$.   All higher-order Green functions will emerge from
time derivatives of composite operators.  This suggests that
we should start with the Hubbard operators, 
$\eta_{\sigma i}=c_{i\sigma}n_{i-\sigma}$ and 
$\xi_\sigma(i)=c_{i\sigma}(1-n_{i-\sigma})$ introduced previously.
The utility of these operators is immediate. Their commutator
with the interaction part of the Hubbard model yields $-(\mu-U)\eta_{i\sigma}$ and
$-\mu\xi_{i\sigma}$, respectively.  Hence, they can be used
to diagonalise the
$t=0$ Hubbard model. Consequently, if this basis is used to compute correlation 
functions, all higher-order correlation functions that will be generated
will be multiplied by the hopping term.  That is, these operators form
the basis for a $t/U\ll 1$ or equivalently a strong coupling expansion. 
The essence of this approach was first proposed by Linderberg and \"Ohrn\cite{ohrn}
and has been applied to the Hubbard model\cite{be,roth,compo} as
 well as the p-d
model for the cuprates\cite{jap}.

To illustrate the utility of the composite operator technique,
we switch to the 4-component basis,
\beq\label{4com}
\psi(i)=\left(\begin{array}{l}
\xi_{i\uparrow}\\\eta_{i\uparrow}\\\xi^\dagger_{i\downarrow}\\\eta^\dagger_{i\downarrow}
\end{array}\right).
\eeq
Let us define 
\beq\label{current}
j(i)=i\partial_t\psi(i)=[\psi(i),H]
\eeq 
as the `current'
operator.  Formally, we can write
the n$^{\rm th}$ element of the current as 
\beq
j_n(i)=\sum_{m}K_{nm}\psi_m(i)+\delta  j_n(i).
\eeq
We can project out the part
of the correction, $\delta j$, that is `orthogonal' to 
the composite operator basis by rewriting
\beq
\delta j_n=\sum_{m l}\langle\{ \delta j_n,\psi_m^\dagger\}\rangle I^{-1}_{ml}
\psi_l 
+\delta\phi_n,
\eeq
where $\delta\phi$ satisfies the equation,
$\{\delta\phi,\Psi^\dagger\}=0$. Let us introduce the normalization matrix
\beq
I_{il} = \langle\{\psi_i,\psi_l^{\dagger}\}\rangle = 
\frac{\Omega}{(2\pi)^2}\int d^2k e^{i\vec{k}\cdot(\vec r_i - \bf r_l)}
 I(\vec{k})
\eeq 
and the overlap matrix
\beq\label{M}
M_{il} = \langle\{j_i,\psi_l^{\dagger}\}\rangle = \frac{\Omega}{(2\pi)^2}
\int d^2k e^{i\vec{k}\cdot(\vec{r}_i - \vec{r}_l)} M_{il}(\vec{k})
\eeq   
where k-integration is over the Brillouin zone and $\Omega$ the
inverse area of the Brillouin zone. 
The elements of the ${\bf I}$ and ${\bf K}$ matrices
contain only the mutual
correlations among the constituents of the composite operator basis.
Because $\{\delta\phi,\Psi^\dagger\}=0$, we can write the overlap matrix
as 
\beq
{\bf M}={\bf E I}
\eeq
where 
\beq
E_{nm}=K_{nm}+\sum_l\langle\{\delta j_n,\psi^\dagger_l\}\rangle I^{-1}_{lm}.
\eeq  
This matrix (the energy matrix)
will play a crucial role in our approach.  Because
of the
$\delta j$ contribution, the energy matrix contains higher order correlations
which are not expressible as simple correlations of the
$\psi$-fields.  It is from the $\delta j$ terms that
the composite excitations arise.
Consequently, the higher-order correlation
functions must either be decoupled or used as self-consistent
parameters in equations obtained by imposing
various symmetries, such as the Pauli principle\cite{mmc,compo}.
As a result of the orthogonality of $\delta\phi$ to the composite
operator basis, we will neglect the contribution
from $\delta\phi$ in all subsequent calculations. This approximation
is equivalent to ignoring the dynamical corrections to
the self-energy.  Such contributions are expected to be
small because the Hubbard operator basis solves
exactly the $U=\infty$ limit.  The principal role
of $\delta\phi$ is to broaden the energy levels associated
with the composite operator spectrum.  The role
of such dynamical corrections will be the focus of 
future study. Within this approximation,
we can write any Green function
\beq
S(i,j)=\langle\langle\psi(i);\psi^\dagger(j)\rangle\rangle
\eeq
in Fourier space as
\beq\label{green}
S({\bf k},\omega)=(\omega-{\bf E(k)})^{-1}{\bf I}({\bf k})
\eeq
by using the equations of motion.  This equation and
the expression, ${\bf M=EI}$, are the central equations of this approach. 
The M and I matrices involve correlation functions which can be 
obtained self-consistently and by introducing
symmetry properties such as the Pauli principle\cite{compo} to close the equations
of motion.  It is the off-diagonal blocks of the energy matrix that contain information about
anomalous pairing correlations.  

There are three types of self-consistent equations that enter this approach.
The simplest of these involves the correlations between the 
$\psi$ fields that appear in $\vec M$ and $\vec I$.
The general expression for such correlation functions in terms of the corresponding Green function is given by
\beq\label{corr1}
\langle\psi_m(i)\psi^\dagger_n(j)\rangle&=&\frac{\Omega}{(2\pi)^2}\int d^2k d\omega
e^{i\vec k\cdot(\vec r_i-\vec r_j)}(1-f(\omega))\nonumber\\
&&\times\left(\frac{-1}{\pi}\right)
{\rm Im} S_{mn}(\bf k,\omega)\nonumber\\
&\equiv&C_{m,n}(i,j)
\eeq
with $f(\omega)$ the Fermi-Dirac distribution function.
The self-consistency
of Eq. (\ref{corr1}) follows from the dependence of ${\bf E}$ and ${\bf I}$
on correlations in the composite operator basis.  To obtain explicitly
the self-consistent equations, we rewrite Eq. (\ref{green}) as
\beq\label{seq}
S({\bf k},\omega)=\sum_{i=1,2}\left[\frac{\kappa_i^+}{\omega-\epsilon_i+i\eta}
+\frac{\kappa_i^-}{\omega+\epsilon_i-i\eta}\right]
\eeq
where
\beq
\kappa_i^\pm=\frac{\lambda(\pm\epsilon_i)}{\pm 2\epsilon_i(\epsilon_i^2-\epsilon_j^2)}
\eeq
with $(i,j)$ chosen from $(1,2)$ but $i\ne j$, $\epsilon_i$ the eigenvalues
of the energy matrix and the matrix $\lambda$ is given by
\beq
\lambda(\omega)={\rm Det}(\omega {\bf 1}-{\bf E})(\omega {\bf 1}-{\bf E})^{-1}{\bf I}.
\eeq
The four eigenvalues correspond to the four composite operator bands:
i=1,2 refer to the $\xi$ and $\eta$ bands, respectively, whereas 
$+,-$ index the particle and hole states, respectively.  If we
now use Eq. (\ref{seq}) in Eq.(\ref{corr1}), we obtain the general self-consistent
equation,
\beq\label{sce}
C_{m,n}(i,l)&=&\frac{\Omega}{2(2\pi)^2}\int d^2k 
e^{i\vec{k}\cdot (\vec{r}_i - \vec{r}_l)}\left[I_{mn}+
\right.\nonumber\\
&&
\left.\sum_{i=1,2}\left(
(\kappa_i^+)_{mn}-(\kappa_i^-)_{mn}\right)\right]\tanh\frac{\beta \epsilon_i}{2}
\eeq
for correlation functions within the composite operator basis
where $T=1/k_B\beta$. 

It is well-known that in approximation schemes of this sort, certain correlation functions which
vanish explicitly as a result of the Pauli principle, self-consistently iterate
to a non-zero value.  To avoid this shortcoming, we explicitly maintained
the symmetries\cite{mmc,compo} imposed by the Pauli principle, namely
\beq\label{pauli}
C_{1,2}(i,i)=\langle\xi_\sigma(i)\eta^\dagger_\sigma(i)\rangle=0.
\eeq
This is the second type of self-consistent integral equation that will be used
in this method.
Likewise, $C_{2,1}(i,i)=0$. As a result of imposing these symmetry relations,
we will find that our theory is completely particle-hole symmetric
about half-filling. 

The final self-consistent equations arise from decoupling
correlations of composite fields that are not contained in the basis described by
Eq. (\ref{4com}).  This procedure will be described in detail
in the next section.  

\section{Computational Machinery}

\subsection{Equations of Motion}

We start first by constructing the
algebra 
\beq
\{\xi_i,\xi_j^\dagger\}=\delta_{ij}\left({\bf 1}+\frac12\sigma^\mu n_\mu\right)\\
\{\eta_i,\eta_j^\dagger\}=-\delta_{ij}\frac12\sigma^\mu n_\mu\\
\{\xi_i,\xi_j\}=\{\eta_i,\eta_j\}=\{\xi_i,\eta^\dagger_j\}=0
\eeq
of the composite operators, with $\sigma^\mu$ for $\mu=1,2,3$
 are the Pauli matrices, $\sigma^0$ is the identity matrix,
and 
\beq
-\frac12\sigma^\mu n_\mu =\left(\begin{array}{cc}
n_{i\downarrow}& -c_{i\downarrow}^\dagger c_{i\uparrow}\\ 
-c_{i\uparrow}^\dagger c_{i\downarrow}&n_{i\uparrow}\\
\end{array}
\right).
\eeq
From these equations, it follows that the I-matrix is given by
\beq
I=\left(\begin{array}{cccc}
1-\frac{n}{2}& 0&0& \Delta_0\\ 
0&\frac{n}{2}&-\Delta_0&0\\
0&-\Delta_0^*&1-\frac{n}{2}&0\\
\Delta_0^*&0&0&\frac{n}{2}
\end{array}
\right)
 \eeq
where $\Delta_0=\langle c_{i\uparrow}c_{i\downarrow}\rangle$
and we have 
assumed that
$\langle n_\uparrow\rangle=\langle n_\downarrow\rangle=n/2$. 
We now turn to the ``current''.  We can construct the ``current''
from the equations of motion
\beq\label{EqM}
j(i) = i\frac{\partial}{\partial t} \psi(i) = [\psi(i),H]
\eeq
for the $\psi$-fields with
\begin{eqnarray}
j_{1}(i) &=& -\mu\xi_{i\uparrow} - \sum_{j} t_{ij}c_{j\uparrow} - 4t\pi_{i\uparrow} \\
j_{2}(i) &=& -(\mu-U)\eta_{i\uparrow} + 4t\pi_{i\uparrow} \\
j_{3}(i) &=& \mu\xi_{i\downarrow}^{\dagger} + \sum_{j} t_{ij}c_{j\downarrow}^{\dagger} + 4t\pi_{i\downarrow}^{\dagger} \\
j_{4}(i) &=& (\mu-U)\eta_{i\downarrow}^{\dagger} - 4t\pi_{i\downarrow}^{\dagger} 
\end{eqnarray}
and
\beq\label{Pi}
 4t\left(\begin{array}{c}\pi_{i\uparrow} \\
 \pi_{i\downarrow}\end{array}\right) = 
\sum_{j}t_{ij}\left(\begin{array}{c} -n_{i\downarrow}c_{j\uparrow} 
+ c_{i\downarrow}^{\dagger}c_{i\uparrow}c_{j\downarrow} -
 c_{i\uparrow}c_{i\downarrow}c_{j\downarrow}^{\dagger} 
\\ -n_{i\uparrow}c_{j\downarrow} + 
c_{i\uparrow}^{\dagger}c_{i\downarrow}c_{j\uparrow} +
 c_{i\uparrow}c_{i\downarrow}c_{j\uparrow}^{\dagger}\end{array}\right). 
 \nonumber
\eeq 
We also need the M-matrix.  The distinct elements of this matrix are
\begin{eqnarray}
M_{11}(\vec{k}) = FT\langle\{j_1,\xi_{\uparrow}^{\dagger}\}\rangle   &=& 
-\mu(1-\frac{n}{2}) - 4te\nonumber\\&&- 4t\alpha(\vec{k})(1-n+p) \nonumber \\
M_{12}(\vec{k}) = FT\langle\{j_1,\eta_{\uparrow}^{\dagger}\}\rangle   &=& 4te - 4t\alpha(\vec{k})(\frac{n}{2}-p)\nonumber  \\
M_{22}(\vec{k}) = FT\langle\{j_2,\eta_{\uparrow}^{\dagger}\}\rangle   &=& -(\mu-U)\frac{n}{2} - 4te - 4t\alpha p  \nonumber \\
M_{13}(\vec{k}) = FT\langle\{j_1,\xi_{\downarrow}\}\rangle   &=& -4t\gamma(\vec{k})\theta + 8t\alpha\Delta_0 - 4t\Delta_{c\xi} \\
M_{14}(\vec{k}) = FT\langle\{j_1,\eta_{\downarrow}\}\rangle   &=&
 -(\mu + 4t\alpha(\vec{k}))\Delta_0\nonumber\\&&+4t\gamma(\vec{k}))\theta - 4t\Delta_{c\eta}\nonumber  \\
M_{23}(\vec{k}) = FT\langle\{j_2,\xi_{\downarrow}\}\rangle   &=&
 (\mu - U - 4t\alpha(\vec{k}))\Delta_0\nonumber\\&& + 4t\gamma(\vec{k}))\theta + 4t\Delta_{c\xi}\nonumber  \\
M_{24}(\vec{k}) = FT\langle\{j_2,\eta_{\downarrow}\}\rangle   &=& -4t\gamma(\vec{k})\theta + 4t\Delta_{c\eta}\nonumber 
\end{eqnarray}
with the normal correlations
\beq\label{e}
e = \langle\xi_i^{\alpha}\xi_i^{\dagger}\rangle - 
\langle\eta_i^{\alpha}\eta_i^{\dagger}\rangle,
\eeq 
and
\beq\label{p}
p = \langle n_{i\sigma}n_{i\sigma}^{\alpha}\rangle + 
\langle c_{i\uparrow}^{\dagger}c_{i\downarrow}
(c_{i\downarrow}^{\dagger}c_{i\uparrow})^{\alpha}\rangle - 
\langle c_{i\uparrow}c_{i\downarrow}(c_{i\downarrow}^{\dagger}
c^{\dagger}_{i\uparrow})^{\alpha}\rangle,
\eeq 
and anomalous correlations
\beq\label{Dcxi}
\Delta_{c\xi} = \langle c_{i\uparrow}\xi_{i\downarrow}^{\alpha}
\rangle - \langle c_{i\downarrow}\xi_{i\uparrow}^{\alpha}\rangle,
\eeq
\beq\label{Dceta}
\Delta_{c\eta} = \langle c_{i\uparrow}\eta_{i\downarrow}^{\alpha}
\rangle - \langle c_{i\downarrow}\eta_{i\uparrow}^{\alpha}\rangle,
\eeq
and
\beq\label{Th}
\theta = \langle c_{i\uparrow}c_{i\downarrow}[n_{\uparrow}
(\vec{r}_i+\hat{x}) + n_{\downarrow}(\vec{r}_i+\hat{x})]\rangle.
\eeq
In these equations, $FT$ signifies the Fourier transform as defined in
Eq. (\ref{M}), $\alpha$ represents an average over nearest-neighbour sites,
$\hat{x}$ indexes nearest-neighbour sites, and
 $\alpha(\vec k)=\cos k_x+\cos k_y$.  Unlike $\alpha(\vec k)$
which is the coefficient of the normal correlation functions, the Fourier
coefficient
$\gamma(\vec k)$ is a coefficient of an anomalous correlation function.  Hence,
it is sensitive to the sign change of the anomalous correlation functions
as $\exp(i\vec k\cdot \vec r)$ is summed over nearest neighbour sites.  For s-wave
symmetry, $\alpha(\vec k)=\gamma(\vec k)$ while for d-wave symmetry
$\gamma(\vec k)=\cos k_x-\cos k_y$.  
The symmetry relationships among the elements of $\vec M$ are as follows:
\beq\label{sym}
M_{11}&=&-M_{33}\nonumber\\
M_{12}&=&-M_{34}\nonumber\\
M_{22}&=&-M_{44}\\
M_{14}&=&-M^*_{32}\nonumber.
\eeq
 The $M_{13}$, $M_{14}$, $M_{23}$, and $M_{24}$ contain
the anomalous correlation functions in particular the $\theta$ correlation
discussed in the introduction.

With the $\vec M$ and $\vec I$-matrices in hand, we now construct the energy matrix.
From the structure of the normalization matrix and the symmetry properties
of the $M$ matrix (Eq. (\ref{sym}), it follows that the
energy matrix is of the form,
\beq
\vec E=\left(\begin{array}{lr}
\vec A& \vec B\\ 
\vec B&-\vec A\\
\end{array}
\right),
 \eeq
where $\vec A$ and $\vec B$ are $2\times 2$ matrices.  
The off-diagonal blocks of the energy matrix determine the gap in the energy
spectrum induced by pairing.  To isolate the relevant parts of the normalization
and $\vec M$ matrices that enter the energy gap, we consider 
the candidate pairing
symmetries.  No simplifications occur in the s-channel.  However, we
have verified that none of the anomalous correlation functions 
are non-zero either
in this channel.  The same state of affairs occurs in p-wave symmetry.
What about d-wave symmetry?  Here several significant simplifications occur. 
First, as a result of the nodes in the gap, no purely on-site anomalous
correlation functions survive.  As a consequence, 
$\langle c_{i\uparrow}c_{i\downarrow}\rangle=\Delta_0=0$.  This leads to a vanishing of
all the off-diagonal blocks of the normalization matrix.  Consequently,
only the off-diagonal blocks of the $\vec M$-matrix enter the off-diagonal
elements of the energy matrix.  However, further simplifications occur.
Consider for example, $\Delta_{c\xi}$ and $\Delta_{c\eta}$ in Eqs. (\ref{Dcxi})
and (\ref{Dceta}).  These quantities are symmetrized and
in addition involve a sum over the nearest neighbour
sites in the Brillouin zone.  Consequently, they vanish identically in
the d-wave channel.  The only anomalous correlation function that remains is $\theta$,
Eq. (\ref{Th}).  At this level of theory, this is the only correlation function
that does not vanish by symmetry conditions.  This is the only anomalous correlation
function that enters the gap in the energy spectrum.  Hence, as advertised,
any pairing will necessarily be governed by a non-traditional 
correlation function.
This correlation function involves a composite operator
that lives on a cluster of nearest-neighbour sites.  By defining
$\eta_{ij\sigma}$ to be $c_{i\sigma}n_{j\tau}$, we rewrite
Eq. (\ref{Th}) as Eq. (\ref{thc}) in which it is clear
that pairing in $\theta$ occurs between two composite excitations.

\subsection{Computational Procedure}

What remains to be done is the calculation of the anomalous
correlation functions.  Our assumption of singlet pairing implies that
\beq
\langle c_{i\uparrow}c_{i\downarrow}n_{j\tau}\rangle=\langle c_{i\uparrow}c_{i\downarrow}n_{j-\tau}\rangle=\langle c^\dagger_{i\downarrow}c^\dagger_{i\uparrow}n_{j\tau}\rangle^*.
\eeq
Consequently, we can define $\theta$ as 
\beq
\theta=2\langle c_{i\uparrow}c_{i\downarrow}n_{j\tau}\rangle
\eeq
where the spin $\tau$ is arbitrary.  As this correlation function involves more than two basis operators, it must be decoupled.  We will follow a procedure analogous to that devised by Roth\cite{roth} in her treatment of the strong-coupling limit of the Hubbard model. To implement the Roth
method, we consider the Green function
\beq
H(i,j,t,t')=\langle\langle c_\downarrow^\dagger(i,t);c_\uparrow(i,t')^\dagger
n_\sigma(j,t')\rangle\rangle.
\eeq
Clearly, if this Green function is calculated, $\theta$ can be obtained directly
from an expression analogous to Eq. (\ref{corr1}). We proceed by constructing
the series of Green functions,
\beq
A_\sigma(i,j,k,t,t')&=&\langle\langle \xi_\uparrow(i,t);c_\uparrow(j,t')^\dagger
n_\sigma(k,t')\rangle\rangle\nonumber\\
B_\sigma(i,j,k,t,t')&=&\langle\langle \eta_\uparrow(i,t);c_\uparrow(j,t')^\dagger
n_\sigma(k,t')\rangle\rangle\\
F_\sigma(i,j,k,t,t')&=&\langle\langle \xi^\dagger_\downarrow(i,t);c_\uparrow(j,t')^\dagger n_\sigma(k,t')\rangle\rangle\nonumber\\
G_\sigma(i,j,k,t,t')&=&\langle\langle \eta^\dagger_\downarrow(i,t);c_\uparrow(j,t')^\dagger
n_\sigma(k,t')\rangle\rangle\nonumber
\eeq
in which the creation operator at time $t$ is replaced by each of 
the composite operators
in the 4-component basis. The Green function of interest, $H$, is obtained
by summing $F$ and $G$ and setting $i=j$.  To calculate these quantities,
we use the equations of motion for $\psi$, Eq. (\ref{current}), and in
particular the approximation that $j=i\partial_t\psi\approx E\psi$.  
Consequently, the equations of motion for the Green functions,
\beq
i\partial_t\left(\begin{array}{l}
A_\sigma\\ 
B_\sigma\\
F_\sigma\\
G_\sigma\\
\end{array}
\right)=E(i)\left(\begin{array}{l}
A_\sigma\\ 
B_\sigma\\
F_\sigma\\
G_\sigma\\
\end{array}\right)+i\delta(t-t')\left(\begin{array}{l}
f_{1\sigma}\\ 
f_{2\sigma}\\
f_{3\sigma}\\
f_{4\sigma}\\
\end{array}\right)
\eeq
are directly related to the energy matrix in real
space, $E(i)$ and the term, $f_{n\sigma}=
\langle\{\psi_n(i,t),c_\uparrow(j,t)^\dagger n_\sigma(k,t)\}\rangle$, that arises from the equal-time anticommutation of $\psi$ with the composite operators in the Green function. Note that in general, $f_n$ is a linear combination
 of correlation functions in the composite operator
basis and correlations associated with $A+B$ or $F+G$.
Consequently, we now have a closed set of equations from which $\theta$ can be obtained.  

In the next step, we Fourier transform these equations of motion
so that $t\rightarrow\omega$, $r_k-r_i\rightarrow k_1$, and $r_j-r_i\rightarrow
k_2$.  Noting that $\omega-E=IS^{-1}$, we obtain
\beq
\left(\begin{array}{l}
A_{\sigma}(k_1,k_2,\omega)\\ 
B_{\sigma}(k_1,k_2,\omega)\\
F_{\sigma}(k_1,k_2,\omega)\\
G_{\sigma}(k_1,k_2,\omega)\\
\end{array}\right)=S(k_1+k_2,\omega)I^{-1} \left(\begin{array}{l}
f_{1\sigma}(k_1,k_2)\\ 
f_{2\sigma}(k_1,k_2)\\
f_{3\sigma}(k_1,k_2)\\
f_{4\sigma}(k_1,k_2)\\
\end{array}\right)\nonumber
\eeq
To extract $\theta$ from these equations, we sum $F$ and $G$, integrate over
$k_2$ so that $i=j$ and then using Eq. (\ref{corr1}), we find that
\beq\label{theta1}
\theta=\frac{2\zeta}{\phi+1}
\eeq
where
\beq\label{theta2}
\phi&=&\frac{n^2-4D}{n(2-n)}\\
\zeta&=&\frac{2}{2-n}\left(C_{11}(i,j)+C_{12}(i,j)\right)\left(C_{13}(i,j)
+C_{14}(i,j)\right)\nonumber\\
&&+\frac{2}{n}\left(C_{22}(i,j)+C_{12}(i,j)\right)\left(C_{24}(i,j)+C_{14}(i,j)\right),
\eeq\label{theta3}
and $D$ is the on-site double occupancy
\beq
D=\langle n_{i\uparrow}n_{i\downarrow}\rangle=\frac{n}{2}-C_{22}(i,i).
\eeq
In the expression for $\phi$, $(i,j)$ denote nearest neighbour sites along the x-axis.  We point out that Beenan and Edwards\cite{be} used a simple
factorization scheme for $\theta$.  As these authors indicate\cite{be}, 
there is no unique way of performing this procedure.  In our scheme, there are two distinct ways of decoupling the Green functions that lead to
$\theta$. That is, we could have started with
another Green function
\beq
H'(i,j,t,t,t')=\langle\langle c_{\uparrow}(i,t);c_{\downarrow}(i,t')c^\dagger_\downarrow(j,t')c_\uparrow^\dagger(j,t')\rangle\rangle.
\eeq
Following the procedure outlined above, we arrive at the alternative expression
for $\theta$:
\beq\label{tprime}
\theta=\frac{2\zeta'}{\phi+1}.
\eeq
with 
\beq
\zeta'&=&\frac{2}{2-n}\left(C_{11}(i,j)+C_{12}(i,j)\right)\left(C_{24}(i,j)
+C_{14}(i,j)\right)\nonumber\\
&&+\frac{2}{n}\left(C_{22}(i,j)+C_{12}(i,j)\right)\left(C_{13}(i,j)+C_{14}(i,j)\right).
\eeq
In the following section, we compare the values of $\theta$ obtained by both methods.  

Similarly it is possible to decouple the parameter
$p$ defined in Eq. (\ref{p}).  Because Eq. (\ref{p}) contains only symmetric combinations of the fermionic operators, it can be decoupled uniquely.
From the procedure outlined above, we find that $p$ is given by
\beq
p=\frac{n^2}{4}-\frac{\rho_1+\phi\rho_2^2}{1-\phi^2}-\frac{\rho_1+\rho_2}{1-\phi}-\frac{\rho_3}{1+\phi}
\eeq
where 
\beq
\rho_1&=&\frac{2}{2-n}\left(C_{11}(i,j)+C_{12}(i,j)\right)^2+\frac{2}{n}
\left(C_{22}+C_{12}\right)^2,\\
\rho_2&=&\frac{2}{2-n}\left((C_{13}(i,j)+C_{14}(i,j)\right)^2+\frac{2}{n}
\left(C_{24}+C_{14}\right)^2\\
\eeq
and
\beq
\rho_3=\frac{4}{n(2-n)}\left(C_{11}(i,j)+C_{12}(i,j)\right)\left(C_{22}+C_{12}
\right).
\eeq
In the normal state, $\rho_2=0$ as it contains only anomalous correlations and
$p$ reduces to the result derived by Roth\cite{roth}.  The expressions for 
$\theta$, $p$ and the self-consistent equation for the correlation functions
constitute the working equations of this method.

To summarize, the equations for $\vec M$ and $\vec I$ contain four parameters,
$\mu$, $e$, $p$ and $\theta$ that must be determined self-consistently. From
Eq. (\ref{e}) and the definition of the Hubbard operators, the
self-consistent equations for $e$ and $n$ are 
\beq
e&=&C_{11}(i,j)-C_{22}(i,j)\\
n&=&2(1-C_{11}(i,i)-2C_{12}(i,i)-C_{22}(i,i)). 
\eeq
In our computational procedure, we either used the two previous equations
together with Eqs. (\ref{theta1}) or (\ref{tprime}).  For the fourth
self-consistent equation, we either used $p$ or the equation
that imposes the Pauli principle, Eq. (\ref{pauli}).  

\section{Results}

\begin{figure}
\begin{center}
\epsfig{file=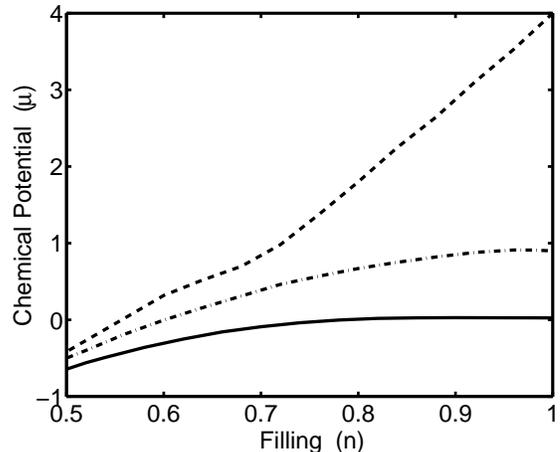, height=6cm}
\caption{The chemical potential (measured in units of the 
hopping t) as a function of filling
for $U=8t$ and $T=0$.  n=1 
corresponds to half filling.  The dashed line corresponds to the 
work of Avella and colleagues\protect\cite{compo} while the dashed
-dotted line to the work of Beenan and Edwards\protect\cite{be}.
Our work is the solid line and includes the effect of pairing.}
\label{fig1}
\end{center}
\end{figure}

\begin{figure}
\begin{center}
\epsfig{file=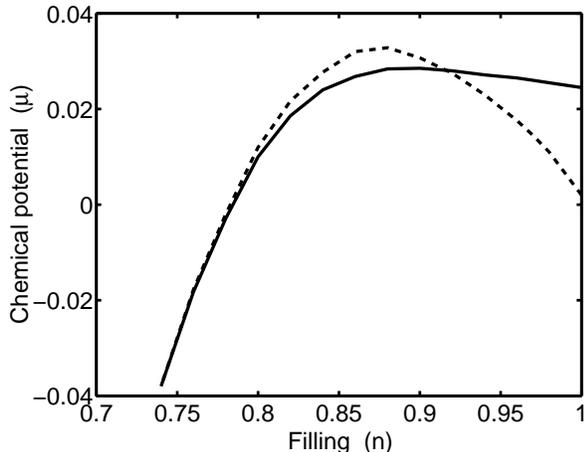, height=6cm}
\caption{The chemical potential as a function of filling 
for $U=8t$ and $T=0$ in the absence of pairing (dashed  line) and in the
 presence
of pairing (solid line).  The chemical potential is in units of $t$.
The solid line clearly illustrates
that pairing alleviates the negative compressibility.}
\label{fig2}
\end{center}
\end{figure}

To set the stage for our results on the anomalous correlations,
we discuss first the chemical potential.  Shown in Fig. (\ref{fig1}) are three 
different
calculations of the chemical potential as a function
of the filling: 1) solid line-- present method,
2) dashed-dotted line--Beenan and Edwards\cite{be},
and 3) dashed line--Avella and co-workers\cite{compo}.
 Here,
$n=1$ corresponds to half-filling.  The method of Avella and 
co-workers\cite{compo} is identical to the method used here.  However, as is
 evident, our results are significantly different.  Particularly striking is 
the continuous increase  
of the chemical potential obtained by Avella and colleagues\cite{compo} as the
 filling is increased. This is in stark contrast our solution
which levels off
in the vicinity of half-filling.  The difference between our 
treatments lies in that there are two solutions to the self-consistent 
equations.  Avella and co-workers chose the solution that is higher in energy 
in the filling
range $0.7-1.0$.  While the lower energy 
solution is the physically-relevant solution, this solution in the absence of 
pairing has a distinct negative compressibility in the vicinity of 
half-filling.  It is for this reason that Avella and co-workers\cite{compo}
 criticized, 
the method used by Roth as it also gives rise to a negative compressibility
as seen from the dashed-dotted line of Beenan and Edwards\cite{be}.
Avella and co-workers\cite{compo} advocated that imposing the Pauli
principle in the self-consistent solution to the equations of motion
eliminates the negative compressibility.
 Our work
shows that even if the Pauli principle is maintained by means of
 Eq. (\ref{pauli}),
a second solution to the integral equations still exists
along which the compressibility remains negative.  To explore further the
 negative compressibility, we show in Fig. (\ref{fig2}) the role of pairing
on the compressibility. 
 The dashed line corresponds to the chemical potential
in the absence of pairing while along the solid line $\theta\ne 0$.  As is
 evident, pairing alleviates the negative compressibility almost entirely
 giving rise to a flattening of the chemical potential in the vicinity of 
half-filling as seen from the solid curve in Fig. (\ref{fig2}).  In fact, a 
key trend common to
the curves shown in Fig. (\ref{fig1}) is that the compressibility is most
positive when pairing vanishes.  
This result is particularly important because a negative compressibility
 occurs in numerous dilute electron systems, such as the 2D electron gas for
 $r_s>3$\cite{2deg}.  Our results also corroborate the earlier observation
of Tandon, Wang, and Kotliar\cite{twk} that the compressibility 
becomes negative in the $U\rightarrow\infty$ Hubbard model.  For
 short-range Coulomb interactions,
 a negative compressibility signifies that the ground state of an electronic 
system is unstable relative to a uniform charge distribution.  Hence, a 
negative compressibility is typically associated with phase separation
or stripe formation\cite{zaanen}.  Our work explicitly shows that pairing
alleviates this instability at least in the case of short-range Coulomb
 interactions.  We have proposed that even in the case of long-range Coulomb
 interactions,
pairing still obtains and alleviates the negative compressibility as 
well\cite{nature}.    

\begin{figure}
\begin{center}
\epsfig{file=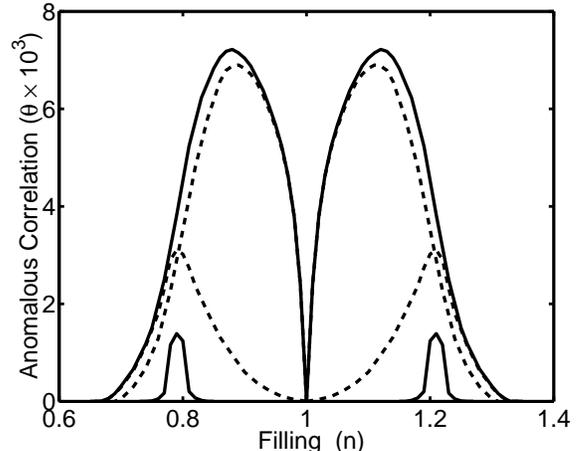, height=6cm}
\caption{Anomalous pairing correlation as a function
of filling for $U=8t$ and $T=0$.  The two solid lines were obtained from 
Eq. (\protect\ref{tprime}).  On the upper curve, the Pauli principle was 
imposed whereas along
the lower curve the decoupling scheme was used to calculate Eq. 
(\protect\ref{p}).  On the dashed curves, Eq. (\protect\ref{theta1}) was
 used to compute
$\theta$.  Once again, along the upper dashed curve, the Pauli principle was 
used whereas along the lower curve the decoupling scheme for $p$
 was used.  Our results
indicate that the anomalous correlations are largest and most stable when
the Pauli principle is imposed.}
\label{fig3}
\end{center}
\end{figure}

We now discuss explicitly the anomalous correlation functions.  Shown in
Fig. (\ref{fig3}) are four different calculations of the correlation
function involving pairing of composite excitations.  
The two solid lines in Fig. (\ref{fig3}) were obtained from 
Eq. (\ref{tprime}).  On the upper curve, the Pauli principle was imposed
whereas along the lower curve, the decoupling scheme was used to calculate 
the parameter $p$ defined in Eq. (\ref{p}).  Hence, the lower curve corresponds
to the method of Beenan and Edwards\cite{be}.
On the dashed curves, Eq. (\ref{theta1}) was used to compute
$\theta$.  Once again, along the upper dashed curve, the Pauli principle was
 used whereas along the lower curve, the decoupling scheme (for $p$) was used.  Our results
indicate that the anomalous correlations are largest and most stable when
the Pauli
principle is imposed by using the integral 
equation, Eq. (\ref{pauli}). 
 When the decoupling scheme is used to
obtain $p$, the two distinct decouplings for $\theta$ yield vastly
different results.  This will result in huge fluctuations in $T_c$ as
the work of Beenan and Edwards\cite{be} illustrates.  What our work establishes
is that if the Pauli principle is imposed
in the self-consistent procedure outlined here, consistent pairing solutions exist
regardless of the decoupling scheme used to calculate $\theta$.  Note also 
that the two lower curves are peaked around a doping level of 20$\%$.  Beenan
and Edwards\cite{be} associated great significance to this doping level as it
corresponds to the filling at which the Fermi surface resembles
the Fermi surface at half-filling for the non-interacting system. 
This appears to
be an accident of their approximations as our more accurate method shows that
the peak in the order parameter occurs at a doping level of 10$\%$.

\begin{figure}
\begin{center}
\epsfig{file=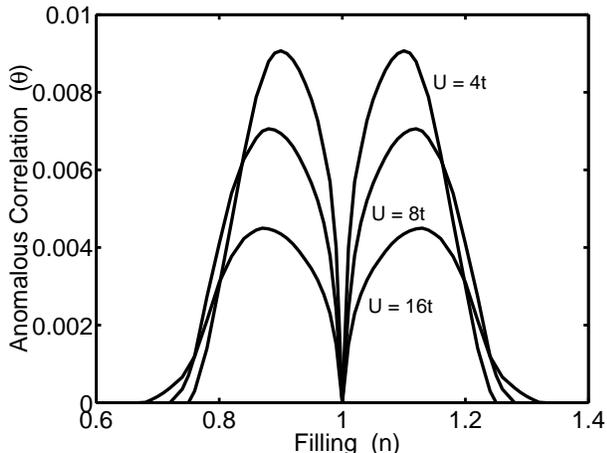, height=6cm}
\caption{Anomalous pairing correlation function as a function
of filling for various values of $U$ for $T=0$.  We obtained
$\theta$
 taking the average of Eqs. (\protect\ref{theta1}) and
(\protect\ref{tprime}).}
\label{fig4}
\end{center}
\end{figure}

\begin{figure}
\begin{center}
\epsfig{file=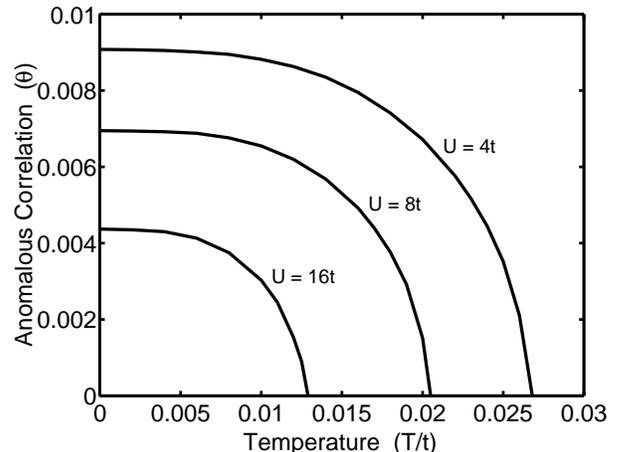, height=6cm}
\caption{Temperature dependence of the anomalous correlation
function at $n=.9$, the optimal
doping as found in Fig. (4).  All quantities are measured
in units of $t$.  The overall shape of $\theta(T)$ is consistent
with the interpretation that $\theta$ is a pairing gap.}
\label{fig5}
\end{center}
\end{figure}

The remaining figures constitute the primary results of this method.
Shown in Fig. (\ref{fig4}) is a calculation of $\theta$ in which the average
of Eq. \ref{theta1}) and (\ref{tprime}) as a function of filling was used. 
Our results
show clearly that $d_{x^2-y^2}$ pairing exists and such pairing is diminished
as $U$ increases. As our method is valid only in the large $U$ limit,
we cannot establish the lower value of $U/t$ for which $\theta$ is 
non-zero.  Also of significance is the maximum in $\theta$ at roughly
10$\%$ doping for the range of on-site repulsions studied. In addition,
the peak in $\theta$ moves to higher dopings as $U$ increases.  We emphasize
that the pair-formation found here does not include 
the effect of phase fluctuations.  Hence, the acutal doping
regime over which true superconductivity with long-range order occurs
might be significantly smaller than that obtained here.  For example, if
long-range order obtains at all, the optimum doping will appear shifted to
higher doping values relative to the maxima of the curves shown here.  
Preliminary results
by Manske, Dahm and Bennemann\cite{mdb} using phenomenological spin-fluctuation
approximations offer some support of this conclusion.
Also of note is the
fact that our treatment preserves the particle-hole symmetry of $\theta$.  
The temperature dependence of the order parameter is shown in Fig. (\ref{fig5}).
As is evident, the shape of $\theta$ is characteristic of any order
parameter that vanishes at a transition temperature.  For $U=8t$,
we find that the onset temperature, $T^\ast=.021t$.  In the cuprates
$t=.5eV$.  Hence, we obtain an onset temperature of $T^\ast=100K$ for pair
formation.  Phase coherence occurs at a lower temperature, $T_c$. 
As a consequence, the $T^\ast$ calculated
here should serve as a realistic estimate of the psuedogap
temperature\cite{psuedo}.  
At optimum doping,
$T^\ast$ should correspond to $T_c$. Experimentally,
optimal doping corresponds to roughly 20\%.  At this doping
level and for $U=8t$, we obtain that $T_c=65K$ leading to a ratio of 
$T^\ast/T_c=1.5$.  For YBa$_2$Cu$_3$O$_{6.95}$, $T_c=92K$ while
$T^\ast=110K$\cite{martindale}. Typically in the cuprates, $1.2<T^\ast/T_c<2$ in
the underdoped regime\cite{psuedo}.  
While our calculated values for $T^\ast$ and $T_c$
are not rigorous estimamtes, it is encouraging
that they are not far off from the experimental values. 

\section{Final Remarks}

We have presented here an analysis ideally suited for
the strong-coupling limit of 
the Hubbard model.  Two key results are established.  First, 
pairing in the Hubbard model occurs in the d$_{x^2-y^2}$ channel.  Second,
 composite
excitations rather than electon-like excitations pair together.  They live on a cluster
of nearest-neighbour sites and are formed out of a hole and a singly-occupied 
site.  Ultimately, such excitations could explain
the absence of a well-defined electron-like peak in the ARPES
experiments\cite{arpes} if the self-energy correction in Eq. (\ref{seq}) arising
from the dynamical processes significantly broadens the energy levels near the Fermi
energy.  Sasaki, Matsumoto and Tachiki\cite{smt} have evaluated
such dynamical corrections in the context of the 
p-d model for the cuprates and found that the composite operator spectrum
remains intact (thereby justifying the initial choice of
the composite operator basis);  however, broadening of all levels including
those at the Fermi surface was observed. Matsumoto and Mancini\cite{matsu}
 have also performed
similar calculations for the Hubbard model\cite{matsu} 
and observed a broadening of the levels
at the Fermi surface. 

Longer-range Coulomb interactions,
necessitate the retension of higher-order 
composite excitations. The simplest of such excitations  will  
involve three sites.  Computing the equations of motion for the
three-site composite excitations leads 4-site excitations.
At each iteration of this procedure, the range of the
composite excitations grows.  We have verified that
nearest-neighbour repulsive Coulomb interactions drastically diminish the anomalous
pairing in $\theta$.  Trivially,
attractive nearest-neighbour Coulomb interactions enhance pairing.
However, to completely settle the issue of pairing
when repulsive next-nearest neighbour interactions
are included, we must also
retain the anomalous correlation functions that are generated
in the presence of the longer-range interaction. 
Our work here suggests that such correlation
functions should be computed to determine if composite particle pairing
survives in the extended Hubbard model.

Nonetheless, within the on-site repulsive model, the correlation
function ($\theta$) calculated here should be sufficient to describe the pairing
process. Because $\theta$ involves
a product of the form $\eta_{ij\uparrow}\eta_{ij\downarrow}$, the 
pairing mechanism is entirely local.  From the 
form of $\eta_{ij}$, it is tempting to conclude that
pairing requires a doubly-occupied site to neighbour a singly-occupied
 site. In such a configuration,
double occupancy can be shared between
the two sites.  However, this is just one of the many types of local
configurations that gives rise to a non-zero value of $\theta$.
If long-range phase coherence obtains, one can think of the condensate as a
coherent superposition of all such resonating structures. This suggests
that the pair-pair correlation function for the cexons should be calculated
to verify one way or another if phase coherence obtains.

$^*$Current Address: Theory Division, Los Alamos National Laboratory.
\acknowledgements
We thank P. Wolynes, A. Yazdani, and T. Leggett for helpful conversations and
the NSF grant No. DMR98-96134.

\end{document}